%% file: root.tex
\documentclass[letterpaper, 10pt, conference,dvipsname]{ieeeconf}

\IEEEoverridecommandlockouts                              

\usepackage{cite}
\let\labelindent\relax
\usepackage{enumitem}

\usepackage{times}
\usepackage{soul}
\usepackage{url}
\usepackage[hidelinks]{hyperref}
\usepackage[utf8]{inputenc}
\usepackage[small]{caption}
\usepackage{graphicx}
\usepackage{booktabs}
\usepackage{algorithm, algpseudocode}
\usepackage[switch]{lineno}
\usepackage{amsmath,bbold,amsthm}
\usepackage{xcolor,nicematrix}

\input{defs}

\usepackage{diagbox}
\usepackage{subfigure}
\usepackage{times,bm}
\usepackage{booktabs}
\usepackage[hidelinks]{hyperref}
\allowdisplaybreaks[2]

\begin{document}
\title{Mitigating Transient Bullwhip Effects Under Imperfect Demand Forecasts}
\author{Sarah H.Q. Li,
Florian D\"{o}rfler 
\thanks{S.H.Q. Li and F. D\"{o}rfler are with the Automatic Control Laboratory, ETH Z\"{u}rich, Physikstrasse 3, Z\"{u}rich, 8092, Switzerland (email: (sarahli@control.ee.ethz.ch, dorfler@ethz.ch)).}%
}
\markboth{Journal of \LaTeX\ Class Files,~Vol.~14, No.~8, August~2015}%
{Shell \MakeLowercase{\textit{et al.}}: Bare Demo of IEEEtran.cls for IEEE Journals}
\newcommand{\urg}[1]{{\color{red} #1}}

\maketitle
\thispagestyle{plain}
\pagestyle{plain}

\begin{abstract}
Motivated by how forecast errors exacerbate order fluctuations in supply chains, we leverage robust feedback controller synthesis to characterize, compute, and minimize the worst-case order fluctuation experienced by an individual supply chain vendor.
Assuming bounded forecast errors and demand fluctuations, we model forecast error and demand fluctuations as inputs to linear inventory dynamics, and use the $\ell_\infty$ gain to define a transient Bullwhip measure. In contrast to the existing Bullwhip measure, the transient Bullwhip measure explicitly depends on the forecast error. This  enables us to separately quantify the transient Bullwhip measure's sensitivity to forecast error and demand fluctuations. To compute the controller that minimizes the worst-case peak gain, we formulate an optimization problem with bilinear matrix inequalities and show that it is equivalent to minimizing a quasi-convex function on a bounded domain. We simulate our model for vendors with non-zero perishable rates and order backlogging rates, and prove that the transient Bullwhip measure can be bounded by a monotonic quasi-convex function whose dependency on the product backlog rate and perishing rate is verified in simulation. 
\end{abstract}
\input{intro}
\input{supply_chain_model}
\input{peak_gain_min}
\input{peak_gain_quasicvx}
\input{sim}
\bibliographystyle{IEEEtran}
\bibliography{reference}
\appendix
\input{app}
\end{document}

%% file: defs.tex
\newcommand{\reals}{{\mathbb{R}}}

\newcommand{\diag}{\mathop{\bf diag}}

\newcommand{\norm}[1]{\left\lVert#1\right\rVert}
\newcommand{\mnorm}[1]{{\left\vert\kern-0.25ex\left\vert\kern-0.25ex\left\vert #1 
    \right\vert\kern-0.25ex\right\vert\kern-0.25ex\right\vert}}

\newcommand{\mc}{\mathcal}

\newcommand{\half}{\frac{1}{2}}

\newtheorem{definition}{Definition} 
\newtheorem{theorem}{Theorem}
\newtheorem{lemma}{Lemma}

\newtheorem{assumption}{Assumption}
\newtheorem{example}{Example}

%% file: intro.tex
\section{Introduction}
As modern supply chains become more interconnected both geographically and between different commodities, an individual disruption's impact on the global supply chain has compounded~\cite{evergreen2021}. Individual supply chain components are connected to a greater number of disruptions worldwide, such that individual disruptions have greater capacity to cause  global supply chain failures~\cite{wagner2006empirical,ranganathan2022re}. 

The amplification of demand fluctuations within a supply chain is termed the {\em Bullwhip effect}~\cite{lee1997bullwhip}. Both behavioral and mathematical analyses point to demand forecast errors as one of the Bullwhip effect's major instigator~\cite{zhang2004impact,chiang2016empirically}. As predictors of future demand, forecasts are unavoidably inaccurate. Their errors induce 1) conservative ordering~\cite{chen2000impact}, when suppliers over-react to demand forecasts by overstocking on supplies; 2) virtual inflation~\cite{wang2016bullwhip}, when consumers suspect potential supply shortages and inflate the demand;  and 3) information withholding~\cite{costantino2014impact}, when competing vendors withhold accurate demand signals from their suppliers in order to retain greater market power.

Despite their role in instigating the Bullwhip effect, forecasts are irreplaceable in  supply chain management. 
Previous work used robust control theory and feedback-based controllers to minimize its effects~\cite{burrows2023who,ouyang2010bullwhip}, yet
most of the existing supply chain analysis under the robust control framework assumes specific forecast tools and therefore cannot generalize to alternative forecasting techniques. 
Presently, we address the following question: \emph{is it possible to model the forecast-driven inventory dynamics as a state-based linear  system, and apply robust control techniques to  mitigate its transient demand amplification?}

We specifically focus on the \emph{transient} demand amplification as opposed to the asymptotic total demand amplification that is used as the standard proxy for the Bullwhip effect in supply chain literature~\cite{ouyang2010bullwhip,udenio2017behavioral}. Historically, most supply and demand trends stabilize in the long term, but the transient fluctuations will still lead to supply chain loss. For example, chicken wings were at first overstocked and then under-supplied at the beginning and middle of the COVID-19 pandemic, respectively~\cite{chicken2021}. While the chicken wing supply chain has restabilized, these fluctuations led to significant and irreversible food wastage. Transient Bullwhip effects have be studied in the frequent domain \cite{dejonckheere2003measuring,wang2016bullwhip} but not in the discrete state space setting we consider in this manuscript.

\textbf{Contributions}. Building on an existing state-space, discrete-time LTI model~\cite{udenio2017behavioral} of a single commodity vendor with non-zero order backlog rates, we incorporate imperfect demand forecast both as a disturbance-driven variable and a feedforward term to augment the state-based feedback control. From this model, we define the transient Bullwhip measure as the worst-case peak gain from forecast errors and demand fluctuations to order fluctuations. We show that the bilinear matrix inequality that bounds the peak gain of a given forecast-driven feedback controller can be reduced to a linear matrix inequality. Furthermore, we show that the transient Bullwhip measure depends nonlinearly on the worst-case magnitudes of the forecasting errors and the demand fluctuations, and empirically compare the impact of different backlog and commodity perishing rates on the transient Bullwhip measure. 

Although we formulate and compute the transient Bullwhip measure for a single vendor, our results can extend to the multi-vendor setting. Mitigating the Bullwhip effect in the multi-vendor setting is more challenging due to the Bullwhip effect's dependence on the supply chain's interaction and communication structure~\cite{burrows2023who,hall2024game,ouyang2010bullwhip}.


%% file: supply_chain_model.tex
\section{Supply chain Inventory Model}\label{sec:model}
We model a single-product supply chain vendor facing a supply-agnostic market and orders and resupplies at fixed time intervals as discrete-time dynamics~\cite{udenio2017behavioral,ouyang2010bullwhip,ivanov2018survey} with the following non-negative ($\reals_+$) variables. 
\begin{enumerate}
    \item \textbf{Order/resupply time interval} $\Delta t\in\reals_+$: a fixed time interval for ordering and receiving, indexed by $k \in \mathbb{N} $, such that $t_{k} = k\Delta t$.
    \item \textbf{Inventory} $i(k) \in \reals_+$: the inventory held at time $t_{k}$.
     \item \textbf{Pipeline} $p(k)\in\reals_+$: the unfulfilled orders at time $t_{k}$.
    \item \textbf{Demand} $d(k) \in\reals_+$: the demand realized at time $t_{k}$.
    \item \textbf{Forecast} $f(k) \in \reals_+$: the forecast made at time $t_{k}$ for the predicted demand at $t_{k+2}$.
    \item \textbf{Previous forecast} $\hat{f}(k) \in \reals_+$: the forecast made at  $t_{k-1}$ for the demand at $t_{k+1}$.
    \item \textbf{Order} $o(k) \in \reals_+$: the order placed at time $t_{k}$.
    \item \textbf{Perish rate} $\beta \in [0,1]$: the percentage of the stocked inventory that expires during each time interval.
    \item \textbf{Backlog rate} $\alpha \in [0, 1]$: the percentage of unfulfilled order at every resupply time interval $\Delta t$.
\end{enumerate}

\noindent\textbf{Inventory dynamics}. During the interval $[t_{k-1}, t_{k}]$, the vendor sells $d(k)$ amount of inventory, receives $(1-\alpha) p(k-1)$ amount of supplied inventory from its pipeline, and retains $(1 - \beta)i(k-1)$ amount of previous inventory. We do not model the shelf life and the time spent in pipeline of individual products. Mathematically, this implies that the  pipeline $p(k)$ and the inventory $i(k)$ are memory-less: at time $t_k$, $(1 - \alpha)$ of the pipeline $p(k-1)$ transfers from the pipeline to the inventory, $\alpha $ of the pipeline $p(k-1)$ remains to be fulfilled, and $o(k-1)$ amount of order is added to $p(k)$. 



\noindent\textbf{Forecast-driven affine control}. We assume that the vendor can access a forecast oracle $f(k)$ that predicts the demand at time $t_{k+2}$. 
Given the current forecast $f(k)$, the previous time step's forecast $\hat{f}(k)$, the inventory level $i(k)$, and the pipeline level $p(k)$, the vendor adopts an affine control law $g$ to determine the current order $o(k)$, 
\begin{equation}\label{eqn:affine_control}
    \textstyle o(k) = g\big(f(k), \hat{f}(k), i(k), p(k)\big),
\end{equation}
where $g$ is affine in all of its input variables. 
When $f(k)$ is treated as an exogenous signal,~\eqref{eqn:affine_control} is a form of disturbance-state combined-feedback control~\cite{khlebnikov2016control}. 
When $f(k)$ is computed from a model of the demand dynamics,~\eqref{eqn:affine_control} is a form of combined feedback-feedforward control~\cite{zhang2014combined}.  In most LTI descriptions of supply chains, the forecast is modeled as a deterministic function of historical demand~\cite{constantino2013exploring,chen2000impact}.  

Consider the historical average demand given by
\begin{equation}\label{eqn:historical_average_demand}
    \textstyle d^\infty = \lim_{T \rightarrow \infty} \frac{1}{T} \sum_{k=0}^T d(k).
\end{equation}
We consider the class of commodities for which the historical average demand $d^{\infty}$ exists. Furthermore, we assume that the transient demand $d(k)$ is  bounded in distance to $d^\infty$. 
\begin{assumption}[Bounded demand deviation]\label{assum:forecast_value}
There exists $\epsilon_d > 0$, such that the demand $d(k)$ satisfies $|d(k) - d^\infty| \leq  \epsilon_d$ for all $k \in \mathbb{N}$.
\end{assumption}

\noindent\textbf{Bounded forecast error model}.
In supply chains, demand forecast typically utilize temporal aggregation  methods such as ARIMA, ARMAX~\cite{gordon2017price,jin2015forecasting} and machine learning~\cite{carbonneau2008application}. 
We consider  forecasting methods that have bounded errors such as bounded confidence intervals in statistical estimates.
\begin{assumption}[Bounded forecast error]\label{assum:forecast_error}
There exists $\epsilon_{f} > 0$, such that the forecast $f(k)$ satisfies $|d(k+2) - f(k)| \leq \epsilon_f$, for all realized demand $d(k)$, $k \in \mathbb{N}$.
\end{assumption}
Assumptions~\ref{assum:forecast_value} and~\ref{assum:forecast_error} together imply that the forecast $f(k)$'s deviation from the historical average $d^\infty$ is also bounded.  
Combining both assumptions,  the worst-case deviation between $|f(k) - d^\infty|$ is given by
\[\textstyle |f(k) - d^\infty| \leq |f(k) - d(k+2)| + |d(k+2) - d^\infty| \leq \epsilon_d + \epsilon_f. \]
We assume that the affine ordering scheme $g$~\eqref{eqn:affine_control} results in positive orders $o(k)$ and show next how to ensure this. 
\subsection{Steady-state: perfect forecast and stationary demand}
When the forecast perfectly predicts the demand, $f(k) = d(k+2)$ for all $k \in \mathbb{N}$, and when the demand is  stationary, $d(k) = d^{\infty}$ for all $k \in \mathbb{N}$.  Let us take the ordering scheme $g$~\eqref{eqn:affine_control} to be linear in the current inventory, current pipeline, and historical demand,
\begin{equation}\label{eqn:raw_ordering}
   \textstyle o(k) = -\gamma_I i(k) - \gamma_Pp(k)  + \gamma_D f(k) ,
\end{equation}
When the ordering scheme is given by~\eqref{eqn:raw_ordering}, the closed-loop inventory dynamics are given by the following linear time invariant system.  
\begin{equation}\label{eqn:stationary_lti}
    \textstyle\begin{bmatrix}
        i(k+1) \\ p(k+1)
    \end{bmatrix} = \begin{bmatrix}
        1 - \beta & 1 - \alpha \\
        -\gamma_I & \alpha - \gamma_P
    \end{bmatrix}\begin{bmatrix}
        i(k)\\p(k)
    \end{bmatrix} + \begin{bmatrix}
        -1 \\ \gamma_D
    \end{bmatrix}d^\infty.
\end{equation}
We first derive conditions on $\gamma_{I}, \gamma_P,$ and $\gamma_D$ such that the closed-loop dynamics~\eqref{eqn:stationary_lti} is asymptotically stable. 
\begin{lemma}[Steady-state stability]\label{lem:internal_eigenvals}
The inventory dynamics~\eqref{eqn:stationary_lti} is stable if and only if $|\lambda_{+}| < 1$ and $|\lambda_-| < 1$, where $\lambda_{\pm}$ are given by 
\begin{equation}\label{eqn:ev_Ac}
   \textstyle  \lambda_{\pm} = \half \left((a + b) \pm \sqrt{(a - b)^2 - 4\gamma_I(1-\alpha) - 4ab}\right),
\end{equation}
where $a = \gamma_P - \alpha$ and $b = \beta - 1$. 
\end{lemma}
The proof follows by finding for the eigenvalues of the state matrix in~\eqref{eqn:stationary_lti} using the quadratic formula. 
In particular, $\gamma_P = 1 + \alpha - \beta$, $\gamma_I = \frac{(1- \beta)^2}{1 - \alpha}$ ensures that $\lambda_{\pm} = 0$ and achieves the fastest convergence in~\eqref{eqn:stationary_lti} towards steady-state values.
The range of $\gamma_P, \gamma_I$ values that guarantee asymptotic stability matches the empirically verified values from~\cite{udenio2017behavioral}.

If the parameters $(\gamma_I^\infty, \gamma_P^\infty,\gamma_D^\infty)$ satisfy Lemma~\ref{lem:internal_eigenvals} under perfect forecast and stationary demand, they result in the equilibrium state $[i^\infty, p^\infty]^\top \in \reals^2$ that satisfies
\begin{equation}~\label{eqn:equilibrium_state}
   \textstyle\begin{bmatrix}
        i^\infty\\p^\infty
    \end{bmatrix} = \begin{bmatrix}
        1 - \beta & 1 - \alpha  \\
        -\gamma^\infty_I & \alpha  -\gamma^\infty_P
    \end{bmatrix} \begin{bmatrix}
        i^\infty\\ p^\infty
    \end{bmatrix}+  \begin{bmatrix}
        -1 \\  \gamma^\infty_D 
    \end{bmatrix} d^\infty.
\end{equation}
In order for system~\eqref{eqn:stationary_lti} to be a valid description of the supply chain system, the physical quantities $i^\infty, p^\infty,$ and $o^\infty = -\gamma_I^\infty i^\infty - \gamma_P^\infty p^\infty$  need to be strictly positive. 
\begin{assumption}\label{assum:positive_steady_state}
The parameters $(\gamma^\infty_I, \gamma^\infty_P, \gamma^\infty_D)$ produce stable closed-loop dynamics~\eqref{eqn:stationary_lti} and generate strictly positive steady-state inventory, pipeline~\eqref{eqn:equilibrium_state}, and order values~\eqref{eqn:raw_ordering}, i.e.,
   \begin{multline}
     \textstyle \begin{bmatrix}
        i^\infty\\p^\infty
    \end{bmatrix} = \begin{bmatrix}
        \beta & \alpha-1  \\
        \gamma^\infty_I & 1 - \alpha  +\gamma^\infty_P
\end{bmatrix}^{-1}\begin{bmatrix}
        -1 \\ \gamma^\infty_D
    \end{bmatrix}d^\infty > 0, \\
    o^\infty = -\gamma_I^\infty i^\infty - \gamma_P^\infty p^\infty + \gamma_D^\infty d^\infty > 0. 
\end{multline} 
\end{assumption}
\begin{example}[Non-perishable goods]\label{ex:postive_inv}
For non-perishable inventory ($\beta = 0$) and positive historical demand ($d^\infty > 0$), the solution to~\eqref{eqn:equilibrium_state} is given by
\begin{equation*}
 \begin{aligned}
    \textstyle i^\infty& \textstyle = (\gamma_D^\infty - 1 - \frac{1}{1 - \alpha}\gamma_P^\infty)  d^\infty/\gamma_I^\infty, \ p^\infty = d^\infty/(1 - \alpha), \\ 
    \textstyle o^\infty& = -\gamma_I^\infty i^\infty - \gamma_P^\infty p^\infty + \gamma_D^\infty d^\infty.
\end{aligned}   
\end{equation*}
The pipeline value $p^\infty$ and order value $o^\infty$ are independent of the linear control parameters,  and $p^\infty$ is strictly positive if the backlog rate $\alpha < 1$. 
Assuming that $\gamma_I^\infty > 0$, a positive inventory $i^\infty > 0$ requires that  $\gamma_D^\infty > \gamma_P^\infty/(1 -\alpha) + 1$.
In addition, if $\alpha = 0$ (no backlog), then $o^\infty = p^\infty = d^\infty$. 
\end{example}
\subsection{Imperfect forecast and noisy demand}
Returning to a setting in which the forecast $f(k)$ has a forecast error of up to $\epsilon_f$ (Assumption~\ref{assum:forecast_error}) for a demand $d(k)$ whose maximum deviation from the historical average is $\epsilon_d$ (Assumption~\ref{assum:forecast_value}), we offset our inventory dynamics~\eqref{eqn:stationary_lti} to the steady-state values driven by the historical average demand $d^\infty$. 
Given linear control parameters $\gamma_P^\infty, \gamma_I^\infty, \gamma_D^\infty$ that satisfy Assumption~\ref{assum:positive_steady_state}, historical average demand $d^{\infty} \in \reals_{+}$, and corresponding steady-state values $i^\infty, p^\infty, o^\infty \in \reals_+$, we define the shifted state and control variables as
\begin{multline*}
    \textstyle \begin{bmatrix}
        x_1(k)\\x_2(k)
    \end{bmatrix} = \begin{bmatrix}
        i(k)\\ p(k) 
    \end{bmatrix} -  \begin{bmatrix}
        i^\infty\\ p^\infty
    \end{bmatrix}, u(k) = o(k) - o^\infty.
\end{multline*}
The shifted dynamics then satisfy
\begin{equation}\label{eqn:x_1}
  \begin{aligned}
   \textstyle  x_1(k+1) & \textstyle  =  (1 - \beta) x_1(k) + (1 - \alpha) x_2(k)- \big(d(k) - d^\infty\big),\\
   \textstyle  x_2(k+1) &\textstyle  = \alpha x_2(k) + u(k),\\
   \textstyle x_3(k+1) &\textstyle  = f(k-1) - d^\infty,
\end{aligned}  
\end{equation}
where a third state variable $x_3$ is introduced to allow a time delay between the forecast $f(k)$ and the realized demand $d(k+2)$. Under Assumption~\ref{assum:forecast_error}, $d(k) = f(k-2) + \epsilon_1(k)$ for some $|\epsilon_1(k)| \leq \epsilon_f$. Let $w_1(k) = d(k) - f(k-2), w_2(k) = f(k) - d^\infty$, such that $w_1(k) + w_2(k-2) = d(k) - d^\infty$,
then~\eqref{eqn:x_1} is equivalent to 
     \begingroup
    \makeatletter\def\f@size{9.5}\check@mathfonts
\begin{multline}\label{eqn:supply_zeroed_lti}
    \textstyle x(k+1) = A x(k) + B u(k) + B_w \begin{bmatrix}
        w_1(k) \\ w_2(k)
    \end{bmatrix},\\
    \textstyle  A = \begin{bmatrix}
        1 - \beta & 1 - \alpha  & -1\\
        0 & \alpha  & 0\\
        0 & 0 & 0
    \end{bmatrix}, \ B = \begin{bmatrix}
        0 \\ 1 \\ 0
    \end{bmatrix}, \ B_w =  \begin{bmatrix}
        -1 & 0 \\ 0 & 0\\ 0 & 1
    \end{bmatrix},  
\end{multline}  
\endgroup
where we utilize the following forecast-driven  affine control 
\begin{equation}\label{eqn:shifted_control}
     \textstyle u(k) = o(k) - o^\infty =  F_xx(k) + F_w w(k), \,
\end{equation}
for $F_x \in \reals^{1\times 3}, \ F_w = \begin{bmatrix}
        0& g_w
    \end{bmatrix} \in \reals^{1\times 2}, g_w \in \reals$, and $o^\infty = -\gamma_I^\infty i^\infty - \gamma_P^\infty p^\infty + \gamma_D^\infty d^\infty$. We note that the affine control $u(k)$~\eqref{eqn:shifted_control} is not equivalent to the linear control~\eqref{eqn:raw_ordering} with offsets to the state variables. 
    Instead, $o(k)$ under $u(k)$~\eqref{eqn:shifted_control} is affine in the fluctuations in the inventory, pipeline, and forecast values with fixed steady-state offset $o^\infty$, given by
\[\textstyle o(k) = o^\infty + F_x x(k) + F_w w(k),\]
such that the control matrices $F_x$ and $F_w$ are unrelated to $(\gamma_I, \gamma_P, \gamma_D)$.
Under Assumptions~\ref{assum:forecast_value} and~\ref{assum:forecast_error}, the instantaneous disturbance $w(k)$ for any $k$ belongs to the set $\mathcal{W}$, given by
\begin{equation}\label{eqn:forecast_and_demand_flux}
   \textstyle \mathcal{W} = \left\{\begin{bmatrix}
        w_1(k)\\w_2(k)
    \end{bmatrix}\in \reals^2 \ \vline  \ |w_1(k)| \leq \epsilon_f, |w_2(k)|\leq \epsilon_f + \epsilon_d \right\}. 
\end{equation}
The disturbance term $w_2(k)$ correspond to the forecast fluctuation from the historical demand average and is realized at time step $k$. The true demand $d(k+2)$ is realized at time step $k+2$, corresponding to $w_1(k+2)$.

We assume the following to ensure that the supply chain variables~\eqref{eqn:supply_zeroed_lti} remain positive under the disturbance set $\mc{W}$.
\begin{assumption}[Sufficiently small demand variations]\label{assum:stable_origin}
For a chosen set of control parameters $(\gamma^\infty_p, \gamma^\infty_I, \gamma^\infty_d)$ and a given historical demand $d^\infty > 0$ , the disturbance set $\mc{W}$~\eqref{eqn:forecast_and_demand_flux}'s bounds $\epsilon_{d}, \epsilon_{f}$ satisfy $d^\infty, i^\infty, p^\infty, o^\infty \gg \epsilon_d + \epsilon_f$,
where $i^\infty, p^\infty, o^\infty$ are the steady-state values that result from the control parameters $(\gamma^\infty_p, \gamma^\infty_I, \gamma^\infty_d)$~\eqref{eqn:equilibrium_state}. 
\end{assumption}



%% file: peak_gain_min.tex
\section{Transient Bullwhip as peak-to-peak gain}\label{sec:transient_bw}
In~\cite[Def.2]{ouyang2010bullwhip}, the Bullwhip effect is characterized by the worst-case ratio of total order fluctuation to total demand fluctuation over the infinite time horizon.
Adapted for the single vendor dynamics~\eqref{eqn:supply_zeroed_lti} and bounded demand (Assumption~\ref{assum:forecast_value}), the vendor experiences the Bullwhip effect if
\begin{equation}\label{eqn:og_Bullwhip}
    W_E:= \sup_{|w_1(j)|  \leq \epsilon_d, \forall j \in \mathbb{N}} \frac{\sum_{k=0}^\infty (o(k) - o^\infty)^2}{\sum_{k=0}^\infty (d(k) - d^\infty)^2} > 1.
\end{equation}
Intuitively, $W_E$~\eqref{eqn:og_Bullwhip} minimizes the energy transfer  from the demand fluctuations $d(k) - d^\infty$ to the order fluctuations $o(k) - o^\infty$ in their $\ell_2$ norm. As a metric for the Bullwhip effect, $W_E$ fails to explicitly account for demand forecast inaccuracy and does not capture transient order fluctuations. Not explicitly accounting for forecasting error makes sense when a specific forecasting method is used, so that $f(k)$ is a deterministic function of the historical demand $\{d(k-i)\}_{i\in\mathbb{N}}$~\cite{constantino2013exploring}. However, the $W_E$ will conflate the impact of forecast inaccuracy with the impact of demand fluctuations, and cannot generalize to alternative forecasting methods. 
Furthermore, $W_E$ fails to account for transient order fluctuations, as shown in Example~\ref{ex:Bullwhip_measure}.
\begin{example}\label{ex:Bullwhip_measure}
We consider two different order time series, $o_1(k)$ and $o_2(k)$ under the same demand fluctuations $d(k) - d^\infty$ and the same steady-state order level $o^\infty = d^\infty$ as shown in Figure~\ref{fig:order_dists}.  
\begin{figure}[ht!]
   \centering
   \includegraphics[width=0.6\columnwidth]{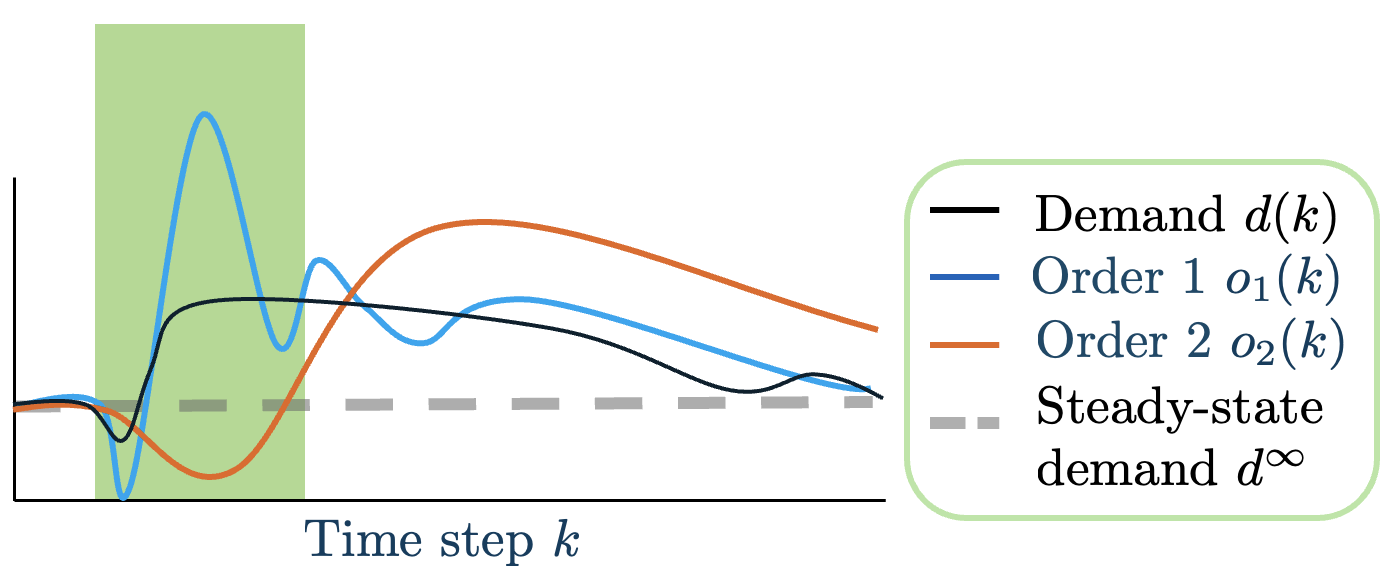}
   \caption{Comparing the transient behavior of two ordering schemes.}
   \label{fig:order_dists}
\end{figure}
Comparing $W_E$~\eqref{eqn:og_Bullwhip} for the order series $o_1$ and $o_2$, we observe that the first order series $o_1(k)$ returns to the steady-state value much faster than the second order series $o_2(k)$, therefore $W_E(o_1) < W_E(o_2)$. However, $\max_{k} |o_1(k) - o^\infty| > \max_k |o_2(k) - o^\infty|$, and during the time period highlighted in green, $|o_1(k) - o^\infty| $ fluctuates significantly more than $|o_2(k) - o^\infty|$. We conclude that $o_1$ creates higher transient virtual demand than $o_1(k)$ despite scoring lower on the metric $W_E$~\eqref{eqn:og_Bullwhip}.
\end{example}
As shown in Example~\ref{ex:Bullwhip_measure}, minimizing the energy-based metric $W_E$~\eqref{eqn:og_Bullwhip} can create higher transient order fluctuations. This is  problematic for supply chain logistics, where a vendor's ability to handle transient supply and demand fluctuations is bound by hard constraints arising from warehouse sizing and transportation availability. Transient order fluctuations can also cause excess stock and perceived shortages which can cascade into chain-wide disruptions~\cite{chicken2021,evergreen2021}. Instead, we propose an alternative metric that evaluates the transient energy transfer from demand fluctuations and forecast error to order fluctuations.
\begin{definition}[Transient Bullwhip Measure]\label{def:Bullwhip_measure}
The inventory dynamics~\eqref{eqn:supply_zeroed_lti} under the forecast-driven affine controller~\eqref{eqn:shifted_control} and demand and forecast fluctuations $\mathcal{W}$~\eqref{eqn:forecast_and_demand_flux} have a transient Bullwhip measure given by 
\begin{equation}\label{eqn:Bullwhip}
W_T = \underset{w(k) \in \mathcal{W}, k\in \mathbb{N}}{\sup} |o(k) - o^\infty|.
\end{equation}
\end{definition}
The transient Bullwhip measure is the $\ell_{\infty}$ gain~\cite{boyd1987comparison} of $u(k)$ from~\eqref{eqn:supply_zeroed_lti}. 
In contrast to the energy-based metric $\mathcal{W}_E$~\eqref{eqn:og_Bullwhip}, $\mathcal{W}_T$~\eqref{eqn:Bullwhip} bounds the worst-case transient behavior of the vendor's order dynamics. When applied to the two different order schemes from Example~\ref{ex:Bullwhip_measure}, the $W_T$ of $o_2$ will be lower than the $W_T$ of $o_1$, thus more accurately reflecting the potential damage caused by order fluctuations and the stress on the supply chain logistics. 

\subsection{Bounding the transient Bullwhip measure}
In general, the transient Bullwhip measure $W_T$~\eqref{eqn:Bullwhip} can be difficult to compute. Instead, we build an upper bound for it using \emph{inescapable ellipsoids}~\cite{abedor1996linear}. 
\begin{definition}[Inescapable ellipsoid]
    For $P \in \reals^{n\times n}$, $P \succ 0$, the set $\Phi(P) = \{x \in \reals^n \ | \ x^\top Px \leq 1\}$ is an inescapable ellipsoid of the dynamics~\eqref{eqn:supply_zeroed_lti} under the state feedback controller $F_x$ and the forecast-driven controller $F_w$ if $x(k) \in \Phi(P)$ for all $w(k) \in \mathcal{W}$ when $x(0) \in \Phi(P)$.
\end{definition}
The inescapable ellipsoid is a type of a robust positively invariant set~\cite{blanchini1999set}, and can be used to bound the worst-case order peak gain achieved by any disturbance $w \in \mc{W}$ and initial state $x(0) \in \Phi(P)$, which is given by
\begin{equation}\label{eqn:ell_bound}
    \max_{x(k) \in \Phi(P), w(k) \in \mathcal{W}}\norm{F_xx(k) + F_w w(k)}_{\infty}.
\end{equation}
All inescapable sets provide an upper bound on $W_T$~\cite{abedor1996linear}. Therefore, we approximate the transient Bullwhip measure via the tightest upper bound available using ellipsoid sets $\Phi(P)$, given by
\begin{equation}\label{eqn:starnorm}
    W_T \leq\inf_{{P} \in \mathcal{S}_{++}^n}\max_{x \in \Phi({P}), w(k) \in {\mathcal{W}}}\norm{F_xx(k) + F_w w(k)}_{\infty}. 
\end{equation}
We first normalize the tightest upper bound~\eqref{eqn:starnorm} to forecast errors and demand deviations in the disturbance set $\mc{W}$~\eqref{eqn:forecast_and_demand_flux}.
\begin{lemma}\label{lem:normalized_gain}
Let the scaled disturbance set~\eqref{eqn:forecast_and_demand_flux} be defined as 
\begin{equation}\label{eqn:scaled_W}
    \textstyle\hat{\mathcal{W}} = \left\{\frac{1}{\hat{\epsilon}}{w} \ | \  w \in \mc{W} \right\},
\end{equation}
where $\hat{\epsilon} = \sqrt{\epsilon_f^2 + (\epsilon_d + \epsilon_f)^2}$.  The transient Bullwhip measure~\eqref{eqn:Bullwhip} is upper-bounded by
    \begin{equation}\label{eqn:normalized_starnorm}
    W_T \leq \hat{\epsilon}\inf_{\hat{P} \in \mathcal{S}_{++}^n}\max_{x \in \Phi(\hat{P}), w(k) \in \hat{\mathcal{W}}}\norm{F_xx(k) + F_w w(k)}_{\infty}. 
\end{equation}
\end{lemma}
The proof is presented in Appendix \ref{app:lem2}.
Recall that checking whether a set $\Phi(P)$ is inescapable can be reduced to evaluating a bilinear matrix inequality. 
\begin{theorem}\label{thm:ellipsoid_bound}~\cite[Thm. 1]{xiao2007peak}
Consider the linear system~\eqref{eqn:supply_zeroed_lti} and the disturbance set $\hat{\mc{W}}$~\eqref{eqn:scaled_W}. The ellipsoid $\Phi(P)$ is inescapable under the controllers $F_x = YP$ and $F_w$ if there exists a scalar value $\lambda \geq 0$ such that 
    \begingroup
    \makeatletter\def\f@size{9.5}\check@mathfonts
    \begin{equation}\label{eqn:peak_gain_bmi}
        \begin{bmatrix}
            -Q + \lambda Q & 0 & QA^\top + Y^{\top}B^\top\\
            0 & -\lambda I & B_w^\top + {F}_w^\top B^\top\\
            AQ + BY & B_w + B{F}_w & -Q
        \end{bmatrix} \preceq 0, Q = P^{-1},  
    \end{equation}
    \endgroup
    where $F_w \in \reals^{1\times 2}$ is the forecast-driven control matrix~\eqref{eqn:shifted_control}.
\end{theorem}
Proof is given in App. \ref{app:thm1}.
Next we show that given $P$, we can compute $\max_{x \in \Phi({P}), w(k) \in \hat{\mathcal{W}}}\norm{F_xx(k) + F_w w(k)}_{\infty}$ via a second linear matrix inequality. Whereas this matrix inequality is typically expressed as a bilinear matrix inequality~\cite{abedor1996linear,xiao2007peak}, the structure of our forecast-driven affine controller enables the bilinear term to be reduced to a linear term, thus resulting in a linear matrix inequality.
\begin{theorem}\label{thm:output_gain_bmi}
Under the state feedback controller $F_x = YP$ and the forecast-driven controller $\frac{1}{\sqrt{\sigma}}F_w$, the $\ell_{\infty}$ gain of the order fluctuations $o(k) - o^\infty$ over the inescapable ellipsoid $\Phi(P)$ and disturbances $\hat{\mathcal{W}}$~\eqref{eqn:scaled_W} is upper bounded as  
\begin{multline}
    \label{eqn:output_gain_thm_statement}
   \max_{x \in \Phi(P), w(k) \in \hat{\mathcal{W}}}\norm{YPx(k) + \frac{1}{ \sqrt{\sigma}}F_w w(k)}_{\infty} \leq \gamma,
\end{multline} 
if and only if there exists a symmetric positive definite matrix $P$, $\lambda \geq 0$, such that~\eqref{eqn:peak_gain_bmi} holds and there exists $\sigma, \gamma > 0$ such that 
    \begin{equation}
        \begin{bmatrix}
            Q & 0 & F_x^\top\\
            0 & (\gamma^2 - \sigma)I & F_w^\top\\
            F_x & F_w &  \sigma I
        \end{bmatrix} \succ 0, \ Q = P^{-1}.
    \end{equation}
\end{theorem}
Proof is given in App \ref{app:thm2}. 
Finding the $\ell_{\infty}$ gain without forecast-driven control involves solving a bilinear matrix inequality~\cite{abedor1996linear}, which the authors reduced to solving an algebraic Riccati equation and a linear matrix inequality. Using the forecast-driven affine control~\eqref{eqn:affine_control}, we avoid solving an algebraic Riccati equation. 

%% file: peak_gain_quasicvx.tex
\section{Bullwhip-minimizing feedback control design}
From Theorems~\ref{thm:ellipsoid_bound} and~\ref{thm:output_gain_bmi},
we can conclude that the forecast-driven affine controller that provides the tightest upper-bound on the transient Bullwhip measure $W_T$~\eqref{eqn:Bullwhip} via inescapable ellipsoids is the optimal argument to the following optimization problem with bilinear matrix inequality constraints.
     \begingroup
    \makeatletter\def\f@size{9.5}\check@mathfonts
\begin{align}\label{eqn:bmi_opt}
     \textstyle W_T \leq& \inf_{\lambda, Q, \sigma,Y, \gamma, F_w} \hat{\epsilon}\gamma\\
        \text{s.t. }& \textstyle \begin{bmatrix}
            -Q + \lambda Q & 0 & QA^\top + YB^\top\\
            0 & -\lambda I & B_w^\top + \hat{F}_w^\top B^\top\\
            AQ + BY & B_w + B\hat{F}_w & -Q
        \end{bmatrix} \preceq 0, \label{eqn:bmi_term}\\
        & \textstyle\begin{bmatrix}
            Q & 0 & Y^\top\\
            0 & (\gamma^2 - \sigma)I & \hat{F}_w^\top\\
            Y & \hat{F}_w &  \sigma I
        \end{bmatrix} \succ 0, \\
         & \textstyle Q \succ 0, \sigma, \gamma > 0, \hat{F}_w = \begin{bmatrix}
             0&F_w&0
         \end{bmatrix}^\top.
 \end{align}
 \endgroup
With optimal values $\gamma^\star$ and optimal arguments $Q^\star, \sigma^\star, Y^\star, F_w^\star$, the peak gain minimizing control is given by  $u(k) = Y^\star(Q^\star)^{-1}x(k) + \frac{1}{\sqrt{\sigma^\star}}F^\star_w w_2(k)$ and the minimal peak gain is given by $\gamma^\star\sqrt{\epsilon_f^2 + (\epsilon_f+ \epsilon_d)^2}$. 

The optimization problem~\eqref{eqn:bmi_opt} is non-convex due the existence of a bilinear matrix term $\lambda Q$ in~\eqref{eqn:bmi_term}. We consider the following function whose point-wise values are semi-definite programs, 
\begin{equation}\label{eqn:alpha_f}
    \begin{aligned}
        f(\lambda) = &\min_{Q,  \sigma,Y, \gamma, F_w}  \gamma^2\\
        \text{s.t. }&  \begin{bmatrix}
            -Q + \lambda Q & 0 & QA^\top + YB^\top\\
            0 & -\lambda I & B_w^\top + F_w^\top B^\top\\
            AQ + BY & B_w + BF_w & -Q
        \end{bmatrix} \preceq 0\\
        & \begin{bmatrix}
            Q & 0 & Y^\top\\
            0 & (\gamma^2 - \sigma)I & F_w^\top\\
            Y & F_w &  \sigma I
        \end{bmatrix} \succ 0, \ Q \succ 0, \sigma, \gamma >  0.
    \end{aligned}
\end{equation}
This function is only well defined on the interval $\lambda \in (0, 1]$.
\begin{lemma}\label{lem:quasiconvex}
The function $f(\lambda)$~\eqref{eqn:alpha_f} not well-defined for $\lambda<  0 $ or $\lambda > 1$. On $\lambda \in (0, 1]$, $f(\lambda)$ is quasi-convex---i.e., for all $\lambda_1, \lambda_2 \in (0,1]$,
\[\displaystyle f(\alpha \lambda_1+(1-\alpha )\lambda_2)<\max {\big \{}f(\lambda_1),f(\lambda_2){\big \}}, \ \forall \alpha \in [0,1].\]
\end{lemma}
Proof is given in App. \ref{app:lem3}.
Given that $f(\lambda)$ is quasi-convex, we can perform a bisection search on $\lambda \in (0, 1]$ to find $\min_{\lambda} f(\lambda)$. However, our simulated results in the following section indicate that the minimum $f(\lambda)$ occurs at $\lim_{\lambda \rightarrow 0} f(\lambda)$, and $f(\lambda)$ is only sensitive to $\lambda$ for non-perishable commodities $\beta \sim 0$. 

%% file: sim.tex
\section{Simulated controller performance}
We evaluate the quasi-convex objective~\eqref{eqn:alpha_f} using CVXPY for different perishing and backlogging rates on the domain $\lambda = [0, 1]$, as well as explore the peak-gain minimizing controller's reliance on the forecast error and its performance under different forecasting methods.

\subsection{Impact of backlog and perishing rates}\label{sec:supchains}
\begin{figure}[ht!]
    \centering
\includegraphics[width=0.9\columnwidth]{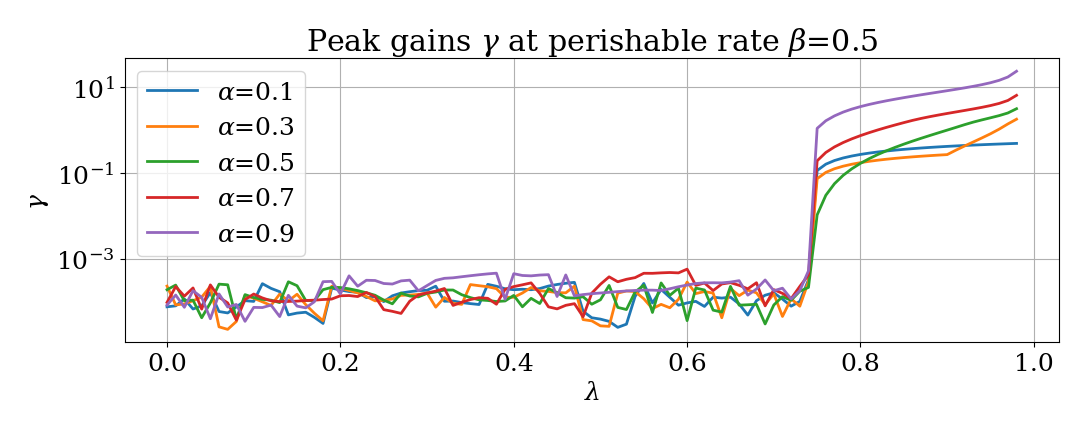}\\
\includegraphics[width=0.9\columnwidth]{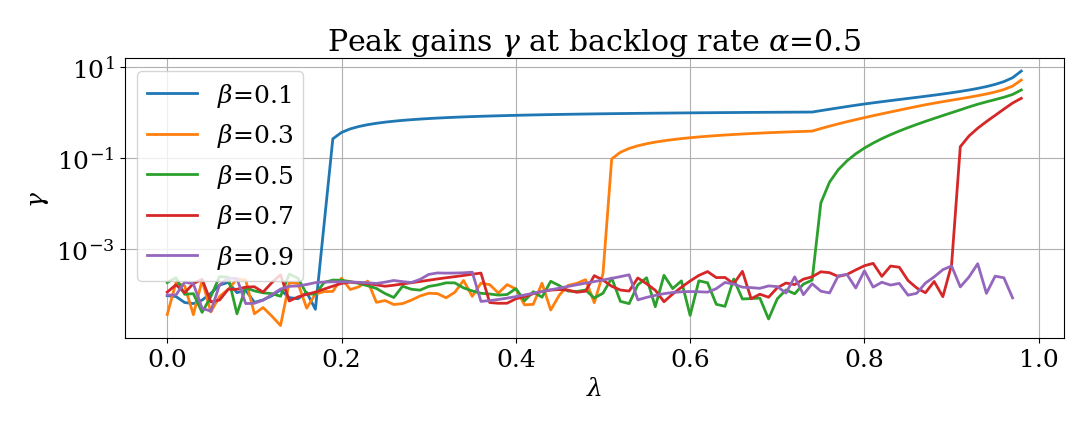}
    \caption{Peak gains $\sqrt{f(\lambda)}$ as a function of $\lambda$ for perishing rate $\beta  = 0.5$ (top) and backlogging rate $\alpha = 0.5$ (bottom) and varying backlog rates.}
    \label{fig:vary_backlog}
\end{figure}

We first let $\hat{\epsilon} = 1$~\eqref{eqn:scaled_W} and evaluate the peak gain $\gamma = \sqrt{f(\lambda)}$~\eqref{eqn:alpha_f} for  different combinations of perishable rates $\beta$ and backlog rates $\alpha$ over $\lambda \in (0,1)$. The results are shown in 
Figures~\ref{fig:vary_backlog} and verify that $f(\lambda)$ does indeed vary with $\lambda$. Specifically,  $f(\lambda)$ appears to be a monotonic function that strictly increases in $\lambda$ for all sampled combinations of backlog and perishing rates. Furthermore, $f(\lambda)$ appears to run into numerical computation errors when $f(\lambda) \sim 1e-8$.

From Figure~\ref{fig:vary_backlog} (top), we observe that the peak gain $\sqrt{f(\lambda)}$ drops significantly around $\lambda \simeq 0.7$ for different backlog rates $\alpha$, while in Figure~\ref{fig:vary_backlog} (bottom), we observe that the same drop in the peak gain $\sqrt{f(\lambda)}$ occurs at different values of $\lambda$, such that the drop occurs at lower $\lambda$ values when $\beta$ increases. Combining these observations together, our simulations show that the backlog rate does not significantly influence the region of $\lambda$ over which the minimum peak gain occurs, while a lower perishing rate $\beta$ causes the minimum peak gain $\gamma^\star$ to be more sensitive $\lambda$, and that in general, evaluating $\sqrt{f(\lambda)}$ at $\lambda$ values close to $0$ could result in better approximations of the minimum peak gain. 
Interestingly, we observe from Figure~\ref{fig:vary_backlog} that neither the backlog nor the perishing rate affects the \emph{minimum} peak gain value $\min_{\lambda} \sqrt{f(\lambda)}$ near $\lambda = 0$. This simulation may imply that all different backlogging and perishing rate combinations have forecast-driven affine controllers that can drive their peak gains very close to zero. 
\subsection{Controller performance under different forecast oracles}
\begin{figure}[ht!]
    \centering
    \includegraphics[width=0.75\columnwidth]{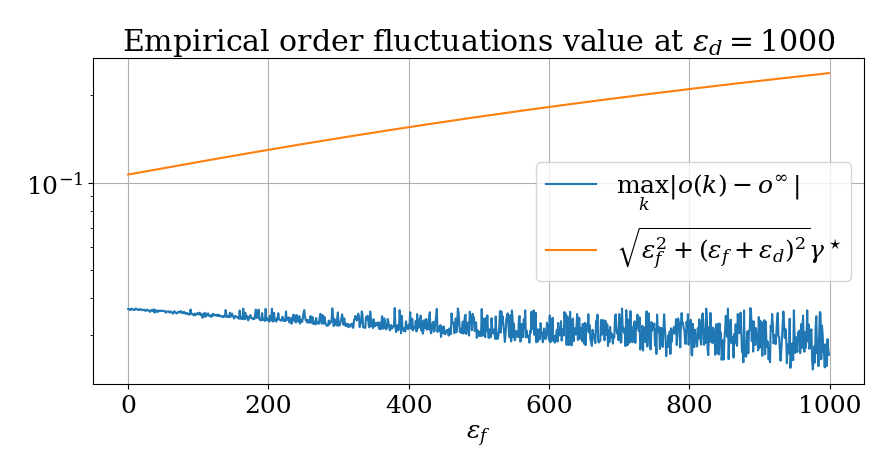}
    \caption{Maximum order fluctuations over $1000$ time steps under increasing forecast error, $\epsilon_f \in [0, \epsilon_d]$. The order fluctuates below $0.04$ and the theoretical bound $\textstyle\gamma^\star \sqrt{\epsilon_f^2 + (\epsilon_f + \epsilon_d)^2}$ monotonically increases between $0.1 - 0.2$.}
    \label{fig:forecast_order}
\end{figure}

\begin{figure}[ht!]
    \centering
    \includegraphics[width=0.75\columnwidth]{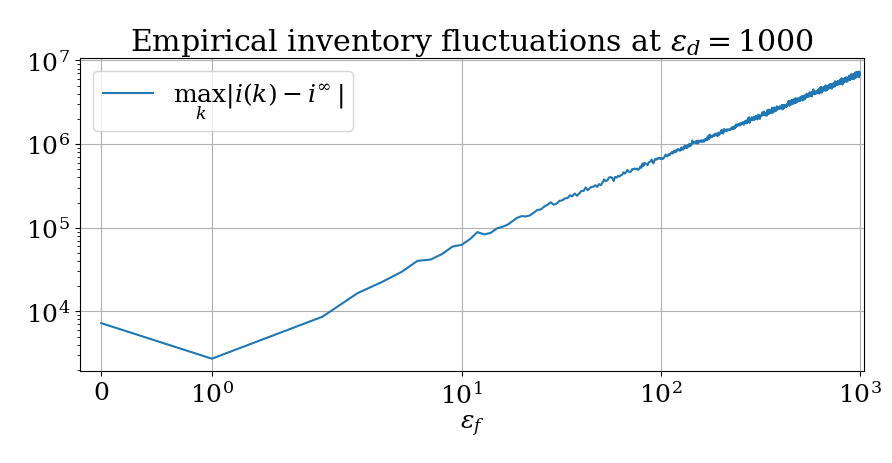}
    \caption{Maximum inventory fluctuations over $1000$ time steps as a function of increasing forecast error, $\epsilon_f \in [0, \epsilon_d]$.}
    \label{fig:forecast_inv}
\end{figure}

Next, we evaluate the peak-minimizing controller's performance for the inventory dynamics~\eqref{eqn:supply_zeroed_lti} with perishing rate $\beta = 0.1$, and backlog rate $\alpha=0.1$. We compute $f(0.01)$~\eqref{eqn:alpha_f}, and use the resulting controllers $F_x^\star$ and $F_w^\star$ for the forecast and demand uncertainty set $\mc{W}$~\eqref{eqn:forecast_and_demand_flux}, with different demand and forecast errors, $\epsilon_d = 1000$ and $\epsilon_f \in [0, 1000]$, respectively. We evaluate the resulting closed-loop dynamics for random initial conditions for $1000$ time steps and observe the maximum fluctuation in order and inventory. The results are shown in Figures~\ref{fig:forecast_order} and~\ref{fig:forecast_inv}. 
We observe from Figure~\ref{fig:forecast_order} that the maximum order fluctuation is well within the range predicted by the predicted transient Bullwhip value, shown in orange. This may imply that our worse-case upper bound is overly conservative. Furthermore, it appears that the maximum empirical order fluctuation is around $0.04$ for all values of $\epsilon_f$. However, we note that we are sampling randomly and not computing the worst-case scenario.    

Interestingly, while forecast error did not appear to impact the empirical order fluctuations, it did significantly impact the empirical inventory fluctuations (shown in Figure~\ref{fig:forecast_inv}, with the inventory fluctuations reaching $10e6$ for $\epsilon_f \simeq 0.2\epsilon_d$, a fairly accurate forecast oracle that is able to reduce the future demand uncertainty five-fold. Finally, we note that the trend for inventory fluctuation is not strictly monotone. At zero forecast error, the inventory fluctuations in fact increase. We repeatedly observed this slight increase at zero forecast error for all repeated trials.  
\section{Conclusion}
We introduce a transient Bullwhip effects using the $\ell_\infty$ gain for a single commodity supply chain vendor driven by forecast uncertainty. The resulting peak gain can be upper-bounded by a quasi-convex function defined on bounded domain. Future work includes extending the model to consider multi-vendor supply chains to further understand the role of individual forecast errors on the  Bullwhip effect. 


%% file: app.tex
\section{Appendix}
\subsection{Proof of Lemma \ref{lem:normalized_gain}}
\label{app:lem2}
\begin{proof}
Let  $x = \hat{\epsilon}\hat{x}$,  ${w} = \hat{\epsilon}\hat{w}$. It follows that 
\begin{enumerate}
    \item $\hat{x}(k+1) = (A +BF_x)\hat{x}(k) + (B_w + BF_w)\hat{w}(k)$ if and only if ${x}(k+1) = (A +BF_x){x}(k) + (B_w + BF_w){w}(k)$;
    \item $w^\top w \leq \hat{\epsilon}^2$ if and only if $\hat{w}^\top\hat{w} \leq 1$;
    \item $x^\top P x \leq 1$ if and only if $\hat{x}^\top \hat{P} \hat{x} \leq 1$ where $\hat{P} = \hat{\epsilon}^2 P$;
    \item $\norm{F_xx(k) + F_w w(k)}_{\infty} = \hat{\epsilon} \norm{F_x \hat{x}(k) + F_w \hat{w}(k)}_{\infty}$. 
\end{enumerate}
We can conclude that the set $\Phi(P)$ is inescapable for the disturbance set $\mc{W}$ if and only if $\Phi(\hat{P})$ is inescapable for the disturbance set $\hat{\mathcal{W}}$. Furthermore, 
\begin{multline}
     \max_{\substack{x(k) \in \Phi(P),\\ w(k) \in \mathcal{W}}}\norm{F_xx(k) + F_w w(k)}_{\infty} = \\
      \hat{\epsilon}\max_{\substack{\hat{x}(k) \in \Phi(\hat{P}),\\ \hat{w}(k) \in \hat{\mathcal{W}}}}\norm{F_x\hat{x}(k) + F_w \hat{w}(k)}_{\infty}.
\end{multline}
Then the problem of finding the minimum upper bound on the transient Bullwhip measure $W_T$ is equivalent to~\eqref{eqn:normalized_starnorm}. 
\end{proof}
\subsection{Proof of Theorem \ref{thm:ellipsoid_bound}}\label{app:thm1}
\begin{proof}
Our results follow from~\cite[Thm.1]{xiao2007peak}, where $\Phi(\hat{P})$ is shown to be inescapable for all disturbance terms $w(k)$ satisfying $w^\top(k) w(k) \leq 1$ if and only if~\eqref{eqn:peak_gain_bmi} holds. Since $\hat{\mc{W}}$~\eqref{eqn:forecast_and_demand_flux} satisfies $w^\top(k) w(k) \leq \epsilon_f^2 + (\epsilon_d + \epsilon_f)^2$, $\Phi(\hat{P}; \hat{\epsilon}^2)$ is inescapable. We note that  $w^\top(k) w(k) \leq \epsilon_f^2 + (\epsilon_d + \epsilon_f)^2$ can be overly conservative depending on the type forecasting method used, but the equality case is realizable in the worst-case scenario under Assumptions~\ref{assum:forecast_error} and~\ref{assum:forecast_value}. 

Next, let $\hat{\epsilon}^2P = \hat{P}$, then $x^\top \hat{P} x \leq \hat{\epsilon}^2$ if and only if $x^\top P x \leq 1$, therefore $\Phi(P; 1) = \Phi(\hat{P}; \hat{\epsilon}^2)$, such that Theorem~\ref{thm:ellipsoid_bound} follows. 
\end{proof}
\subsection{Proof of Theorem \ref{thm:output_gain_bmi}}\label{app:thm2}
\begin{proof}
From~\cite[Lem.4.3]{abedor1996linear}, the two norm of the order peak  under the forecast-driven affine controllers $F_x = YP, \hat{F}_w$ satisfies 
\begin{equation}\label{eqn:pf_0}
    \max_{x(k)^\top P x(k) \leq 1, w(k)^\top w(k) \leq 1}\norm{YPx(k) + \hat{F}_w w(k)}^2_{2} \leq \gamma^2,
\end{equation}
if  and only if there exists $0 < \sigma $, such that 
\begin{equation}\label{eqn:pf_1}
    \textstyle M = \begin{bmatrix}
    \sigma P & 0 & F_x^\top\\
    0 & (\gamma^2 - \sigma)I & \hat{F}_w^\top\\
    F_x & \hat{F}_w & I
\end{bmatrix} \succ 0.
\end{equation}
We substitute $F_x = YP$ and left and right multiply $M$ by $S_1 = \diag\{Q, I, I\}$. Since $S_1 \succ 0$, $S_1MS_1 \succ 0$,
\begin{equation}\label{eqn:pf_sms}
    S_1MS_1 = \begin{bmatrix}
    \sigma Q & 0 & Y^\top\\
    0 & (\gamma^2 - \sigma)I & \hat{F}_w^\top\\
    Y & \hat{F}_w & I
\end{bmatrix} \succ 0.
\end{equation}
From Schur complement,~\eqref{eqn:pf_sms} holds if and only if 
    \begingroup
    \makeatletter\def\f@size{9.5}\check@mathfonts
\begin{equation}\label{eqn:pf_sms2}
   \textstyle S_1MS_1 = \begin{bmatrix}
    (\gamma^2 - \sigma)I & \hat{F}_w^\top\\
    \hat{F}_w & I
\end{bmatrix}  - \begin{bmatrix}
    0 & 0 \\ 0 & \frac{1}{\sigma} YQ^{-1}Y^\top
\end{bmatrix} \succ 0.
\end{equation}
\endgroup
Again, we left and right multiply $S_1MS_1$ by $S_2 = \diag\{I, \sqrt{\sigma} I\}$. Since $S_2 \succ 0$, $S_2S_1MS_1S_2 \succ 0$. 
    \begingroup
    \makeatletter\def\f@size{9.5}\check@mathfonts
\begin{equation}\label{eqn:pf_sms2}
    S_2S_1MS_1S_2 = \begin{bmatrix}
    (\gamma^2 - \sigma)I & \sqrt{\sigma} \hat{F}_w^\top\\
    \sqrt{\sigma} \hat{F}_w & \sigma I
\end{bmatrix}  - \begin{bmatrix}
    0 & 0 \\ 0 & YQ^{-1}Y^\top
\end{bmatrix} \succ 0.
\end{equation}
\endgroup
We use Schur complement again to derive that $S_2S_1MS_1S_2 \succ 0$ if and only if 
\begin{equation}
   \textstyle \begin{bmatrix}
    Q & 0 & Y^\top\\
    0 & (\gamma^2 - \sigma)I & \sqrt{\sigma}\hat{F}_w^\top\\
    Y & \sqrt{\sigma}\hat{F}_w & \sigma I
\end{bmatrix} \succ 0.
\end{equation}
Let $\hat{F}_w =\frac{1}{ \sqrt{\sigma}}F_w$ and noting that
$F_x{x}(k) + \frac{1}{\sqrt{\sigma}}F_w {w}(k)$ is a real number, $\norm{F_x{x}(k) + \frac{1}{\sqrt{\sigma}}F_w {w}(k)}_{2} = \norm{F_x{x}(k) + \frac{1}{\sqrt{\sigma}}F_w {w}(k)}_{\infty}$. 
\end{proof}
\subsection{Proof of Lemma \ref{lem:quasiconvex}}
\label{app:lem3}
\begin{proof}
To see that $\lambda > 1$ and $\lambda < 0$ result in infeasible LMIs, we first note that a necessary condition for an LMI to be negative semi-definite is that its diagonal matrices and principal minors must be negative semi-definite. In particular, this implies that $(\lambda - 1) Q\preceq 0$ and $-Q \preceq 0$. When $\lambda > 1$, $Q$  must be $0$ to satisfy both conditions. 
When $\lambda < 0$, the $(2,2)$ element of the first LMI constraint, $-\lambda I$, cannot be negative semidefinite. 

    
    When $\lambda \in (0, 1]$, let $F_w, Y$ both be the $0$ matrices in their respective dimensions, we can find a set of candidates for the positive semidefinite matrix $Q$ to guarantee the LMI inequality using the Schur complement
    \begin{multline}\label{eqn:last_bmi}
        \textstyle\begin{bmatrix}
        (\lambda - 1)Q & 0 & QA^\top \\
        0 & -\lambda I & B_w^\top \\
        AQ & B_w & - Q
    \end{bmatrix} \preceq 0 \Leftrightarrow \\
   \textstyle -Q -\begin{bmatrix} AQ & B_w\end{bmatrix} \begin{bmatrix}
        \frac{1}{\lambda - 1}Q^{-1} &0\\0 & -\frac{1}{\lambda}I
    \end{bmatrix}\begin{bmatrix} Q A^\top\\ B^\top_w\end{bmatrix} \preceq 0\Leftrightarrow \\
    \textstyle-Q - \frac{1}{\lambda - 1}AQA^\top + \frac{1}{\lambda}B_wB_w^\top \preceq 0.
    \end{multline} 
    Then let  $Q = I + \frac{1}{\lambda} B_wB_w^\top$ and the resulting LMI~\eqref{eqn:last_bmi} will be satisfied. 
    This also holds when $\lambda = 1$, where the corresponding LMI is given by  $-Q  + \frac{1}{\lambda}B_wB_w^\top \preceq 0.$
The result that $f(\lambda)$ is quasi-convex when $f(\lambda)$ is well-defined follows from the proof of~\cite[Thm.3.2]{abedor1996linear}. 
\end{proof}